\documentclass{aa}
\usepackage{graphicx}
\usepackage{natbib}  
\bibpunct{(}{)}{;}{a}{}{,}   

\newcommand{\massp}{m_\mathrm{p}}     
\newcommand{\masse}{m_\mathrm{e}}     
\newcommand{\np}{n_\mathrm{p}}     

\newcommand{\Eeff}{E_\mathrm{eff}}  
    
\newcommand{\Pgas}{P_\mathrm{gas}}    
\newcommand{\Pcr}{P_\mathrm{CR}}    
\newcommand{\Tstart}{t_0}     
\newcommand{\Vmax}{V_{\mathrm{max}}^{\mathrm{ej}}}     
\newcommand{\gamray}{$\gamma$-ray}

\newcommand{\Rej}{R_a}   
\newcommand{\Vej}{V_\mathrm{ej}}   
     
\newcommand{\Bpl}{B_\mathrm{PL}}     
\newcommand{\Bcore}{B_\mathrm{core}}

\newcommand{\Bwd}{B_\mathrm{WD}}     
\newcommand{\BWD}{B_\mathrm{WD}}     
\newcommand{\Rwd}{R_\mathrm{WD}}     
\newcommand{\TP}{test-particle}     
\newcommand{\Vsk}{V_{\mathrm{sk}}}     
\newcommand{\kmps}{km s$^{-1}$}     
\newcommand{\pcc}{cm$^{-3}$}    
\newcommand\Msun{M_{\odot}}     
\newcommand{\xx}[1]{\times 10^{#1}}     
\newcommand\etainj{\eta_\mathrm{inj}}     
\newcommand\etamfp{\eta_\mathrm{mfp}}     
\newcommand\fRad{f_\mathrm{sk}}     
\newcommand\Rsk{R_{\mathrm{sk}}}     
\newcommand\tSNR{t_\mathrm{SNR}}     
\newcommand\EnSN{E_\mathrm{SN}}     
\newcommand\EnCR{E_\mathrm{CR}}     
\newcommand\Mej{M_\mathrm{ej}}     
\newcommand\Bej{B_\mathrm{ej}}     
\newcommand\BISM{B_\mathrm{ISM}}     
\newcommand\rhoISM{\rho_\mathrm{ISM}}     
     
\newcommand\rhoEj{\rho_\mathrm{ej}}     
     
\newcommand\syn{synchrotron}

\newcommand\IC{inverse-Compton}     
     
\newcommand{\rel}{relativistic}     
     
\newcommand\rtot{r_\mathrm{tot}}     
\newcommand\Rtot{r_\mathrm{tot}}

\newcommand\Rsub{r_\mathrm{sub}}     
\newcommand\muG{$\mu$G}     
\newcommand\alf{Alfv\'en}     
\newcommand\Valf{V_\mathrm{A}}     
\newcommand\Msonic{M_\mathrm{S}}     
\newcommand\Malf{M_\mathrm{A}}     
\newcommand\pmax{p_\mathrm{max}}     
\newcommand\Pmax{p_\mathrm{max}}

\newcommand\etal{et al.}     
\newcommand\itt{ }     
\newcommand\bff{ }     
\newcommand{\aaDE}[3]{ 19#1, {\itt A\&A}, {\bff #2}, #3}     
\newcommand{\aatwoDE}[3]{ 20#1, {\itt A\&A}, {\bff #2}, #3}

\newcommand{\apjDE}[3]{ 19#1, {\itt ApJ,} {\bff #2}, #3}     
\newcommand{\apjtwoDE}[3]{ 20#1, {\itt ApJ,} {\bff #2}, #3}     
     
\newcommand{\apjlettwoDE}[3]{ 20#1, {\itt ApJ,} {\bff  #2}, #3}

\newcommand{\appDE}[3]{ 19#1, {\itt Astropart. Phys.,} {\bff #2}, #3}     
\newcommand{\apptwoDE}[3]{ 20#1, {\itt Astropart. Phys.,} {\bff #2}, #3}

\newcommand{\JETPDE}[3]{ 19#1, {\itt JETP, } {\bff #2}, #3}     
     
\newcommand{\mnrasDE}[3]{ 19#1, {\itt M.N.R.A.S.,} {\bff #2}, #3}     
\newcommand{\mnrastwoDE}[3]{ 20#1, {\itt M.N.R.A.S.,} {\bff #2}, #3}

\newcommand{\physrevEDE}[3]{ 19#1, {\itt Phys. Rev. E,} {\bff #2}, #3}

\newcommand{\rppDE}[3]{ 19#1, {\itt Rep. Prog. Phys.,} {\bff #2}, #3}     
     
\newcommand{\ssrDE}[3]{ 19#1, {\itt Space Sci. Rev.,} {\bff #2}, #3}     
\newcommand{\ssrtwoDE}[3]{ 20#1, {\itt Space Sci. Rev.,} {\bff #2}, #3}

\begin{document}     
     
\title{Nonlinear Particle Acceleration at Reverse Shocks in Supernova Remnants}     
     
\author{Donald C. Ellison \inst{1} \and     
        Anne Decourchelle\inst{2} \and
        Jean Ballet\inst{2}
        }

\offprints{D.C. Ellison}

\institute{Department of Physics, North Carolina State 
           University, Box 8202, Raleigh NC 27695, U.S.A.\\
           \email{don\_ellison@ncsu.edu}
       \and 
           Service d'Astrophysique, DSM/DAPNIA, CEA Saclay, 
           91191 Gif-sur-Yvette, France \\
           \email{adecourchelle@cea.fr}     
           \email{jballet@cea.fr}     
           }

\authorrunning{Ellison et al.}

\date{Received July 12, 2004; accepted September 7, 2004}

\abstract{ Without amplification, magnetic fields in expanding ejecta
of young supernova remnants (SNRs) will be orders of magnitude below
those required to shock accelerate thermal electrons, or ions, to
\rel\ energies or to produce radio \syn\ emission at the reverse
shock. The reported observations of such emission give support to the
idea that diffusive shock acceleration (DSA) can amplify magnetic
fields by large factors. Furthermore, the uncertain character of the
amplification process leaves open the possibility that ejecta fields,
while large enough to support radio emission and DSA, may be much
lower than typical interstellar medium values. We show that DSA in
such low reverse shock fields is extremely nonlinear and efficient in
the production of cosmic-ray (CR) ions, although CRs greatly in excess
of $mc^2$ are not produced. 
These nonlinear effects, which occur at the forward shock as well, are
manifested most importantly in shock compression ratios $\gg 4$ and
cause the interaction region between the forward and reverse shocks to
become narrower, denser, and cooler than would be the case if
efficient cosmic-ray production did not occur.
The changes in the SNR structure and
evolution should be clearly observable, if present, and they convey
important information on the nature of DSA and magnetic field
amplification with broad astrophysical implications.

}

\maketitle
     
\section{Introduction}     
     
It is clear in many supernova remnants (SNRs) that the forward, blast
wave shock, interacting with the interstellar medium (ISM) magnetic
field, produces radio (and sometimes X-ray) synchrotron emission.
Presumably this is accomplished when the forward shock accelerates
thermal and pre-existing cosmic-ray (CR) electrons by diffusive shock
acceleration (DSA) (also called the first-order Fermi mechanism).  The
reverse shock, however, will not produce relativistic electrons from
thermal ones if the only magnetic field that is present is the ambient
field from the progenitor star. Any progenitor field will be vastly
diluted by expansion and flux freezing and, for expected white dwarf
or massive star magnetic field values, after $< 100$ yr, will fall
below levels necessary to support particle acceleration to radio
emitting energies.

For example, if the surface field of a white dwarf of radius $\Rwd = 
10^7$ m is $\BWD=10^9$~G, the diluted magnetic field after $\sim 
100$~yr at the reverse shock, which is typically at a radius $\Rsk 
\sim 1$ pc from the explosion site, is $B \sim \BWD (\Rwd/\Rsk)^2 \sim 
10^{-10}$ G.  For $10^{-10}$~G, the diffusive acceleration time to 
10~GeV is $\ga 100$~yr and the upstream diffusion length at a 
$10^4$~\kmps\ shock of a 10 GeV electron is $>1$ pc, making the 
production of radio emitting electrons unlikely. Similar results are 
expected for massive progenitors. Furthermore, with such low fields, 
even if \rel\ electrons are present in large numbers from compressed 
pre-existing cosmic rays or whatever, the radio emissivity may be too 
low to be observable. 

The expanded ejecta bubble may be one of the lowest magnetic field
regions in existence and if reverse shocks in some SNRs are
accelerating electrons by DSA to radio emitting energies or higher 
\citep[as has been suggested by recent radio and X-ray observations,   
e.g.,][]{gotthelf01,delaney02,rho02}, there
are important consequences for:
\begin{itemize}
\item magnetic field generation and amplification in strong     
shocks, 
\item the intrinsic efficiency of DSA and cosmic-ray     
production, including heavy elements, and 
\item the structure and evolution of the X-ray emitting     
interaction region between the forward (FS) and reverse shocks (RS). 
\end{itemize}

At the forward shock in some SNRs \citep[see, in particular, Cas  
A;][]{vink03a}, there is convincing evidence for magnetic fields far
greater than normal ISM values. It has been suggested that the
diffusive shock acceleration process can amplify ambient fields 
\citep[e.g.,][]{lucek00,bell01} to the observed 
levels, and evidence for this is mounting in specific SNRs, 
\citep[e.g.,][]{BKV2003,BPV2003,berezhko04}. If such 
amplification occurs generally, SNRs will be capable of accelerating
cosmic rays to above $10^{17}$ eV \citep[e.g.,][]{ptuskin03,drury2003},
possibly solving the decades old problem of smoothly generating CRs to
the spectral knee near $10^{15}$ eV and beyond. Since DSA is expected
to occur in diverse environments on all astrophysical scales, the
confirmation and characterization of magnetic field amplification is
of extreme importance.
     
In this paper, we first discuss in \S 2 the general effects of a low
magnetic field on the efficiency of DSA.  We show that nonlinear
effects, most noticeable in producing compression ratios $\gg 4$, can
be extremely large for a range of magnetic field below that of the
average ISM (i.e., for $B < \BISM \sim 3\xx{-6}$~G).  We then discuss
in \S 3 the relevance of low fields and strong nonlinear effects in
DSA at reverse shocks in SNRs.  To our knowledge, this is the first
attempt to consider such effects in an evolutionary model of SNRs. A
number of aspects concerning the acceleration process, the nature of
the magnetic field, and the physical conditions in the unshocked
ejecta material (e.g., temperature, ionization fraction, etc.) are not
well-known. Because of these uncertainties, we show a number of
examples where important parameters are varied over fairly wide
ranges.  We emphasize, however, that the effects of efficient DSA on
the structure and evolution of SNRs may be large and current
instruments should be sensitive enough to importantly constrain many
of these poorly known parameters.
     
\section{Nonlinear Diffusive Shock Acceleration}     
     
Nonlinear DSA is a complex process that is difficult to describe
completely. In order to allow the coupling of the particle
acceleration to a hydrodynamic model of SNR evolution, we use an
approximate, algebraic model of DSA developed by \citet{BE99} and
\citet{EBB2000}.
While more complete models exist \citep[e.g.,][]{BEK96}, they tend to be
more computationally intensive and not as easy to include in a
hydrodynamical simulation. Despite the simplifications made in our
acceleration model, we believe it adequately describes the essential
physics when the maximum momentum of accelerated particles $\pmax \gg
A \massp c$.  It is less accurate when $\pmax \sim A \massp c$ as we
describe more fully below. Here, $A$ is the mass number, $\massp$ is the
proton mass, and $c$ is the speed of light.
     
In a complicated, nonlinear fashion, the acceleration efficiency (i.e.,
the fraction of total ram kinetic energy going into superthermal
particles)\footnote{Note the difference between injection efficiency
$\etainj$ and acceleration efficiency. The acceleration efficiency can
and does vary with shock parameters even for a constant $\etainj$.}
depends on the sonic and Alfv\'en Mach numbers
($\Msonic =\sqrt{\rho_0\Vsk^2/(\gamma P_0)}$ and $\Malf =\sqrt{4 \pi \rho_0
\Vsk^2}/B_0$ respectively), on the particle injection and on the  
maximum particle momentum achieved $\pmax$. 
Here, $\rho_0$ is the unshocked mass density,
$P_0$ is the unshocked pressure, $B_0$ is the unshocked magnetic
field, $\Vsk$ is the shock speed, and $\gamma$ is the ratio of
specific heats.
The particle injection is modeled by two parameters: the
injection efficiency $\etainj$, i.e., the fraction of total particles
which end up with superthermal energies, and $\lambda_\mathrm{inj}$,
which determines the value of the injection momentum ($=
\lambda_\mathrm{inj}{A \massp C_\mathrm{s2}}$, where $C_\mathrm{s2}$
is the sound speed in the downstream region).  The parameter
$\lambda_\mathrm{inj}$, by definition $> 1$, is arbitrarily taken to
be 4 in our calculations \citep[see][for a full discussion]{BE99}.

For simplicity, in all our examples we assume a single ion species,
generally protons, but in \S4.4, oxygen, with the electron temperature
equal to the ion temperature $= 10^4$~K.  We ignore any wave damping
from neutral material 
(see \S 5.2.5 below for a discussion).
The maximum momentum protons achieve is
determined by setting the acceleration time equal to the SNR age
$\tSNR$, or by setting the diffusion length of the highest energy
particles equal to some fraction, $\fRad$, of the shock radius $\Rsk$,
whichever gives the lowest $\pmax$ \citep[see, for
example,][]{BaringEtal99}. 

We assume strong turbulence (i.e., Bohm diffusion) so that the
scattering mean free path is on the order of the gyroradius, i.e.,
$\lambda \sim \etamfp r_g$, with $\etamfp= 1$.  The magnetic field
strength is thus an important factor in determining $\pmax$.  If the
turbulence is, in fact, weaker than Bohm diffusion ($\etamfp > 1$),
$\pmax$ will be smaller for a given background $B$ and the shock
compression ratio $\Rtot$ will be
less.
Consistent with assuming strong turbulence ($\etamfp=1$), we set,
following \citet{VBKR2002}, the downstream magnetic field $B_2=
\sqrt{1/3 + 2 \Rtot^2/3}~B_0$.\footnote{Everywhere, the subscript 0
(2) implies unshocked (shocked) quantities.}
    
In the acceleration model of \citet{BE99}, the magnetic field also    
enters in a calculation of the transfer of energy from energetic    
particles to the background gas via \alf\ waves, i.e.,    
\begin{equation}    
\frac{u \rho^\gamma}{\left(\gamma - 1 \right)}     
\frac{\partial}{\partial x}     
\left ( \Pgas \rho^{-\gamma}\right ) =    
\Valf \frac{\partial \Pcr }{\partial x}    
\ ,     
\end{equation}    
where $\Valf = B / \sqrt{4 \pi \rho}$ is the \alf\ velocity, $u$ is the    
flow speed, $\Pgas$ is the pressure in the background gas, and $\Pcr$    
is the pressure in the \rel\ particles, i.e., cosmic rays. It is    
implicitly assumed that the turbulence saturates when $\delta B/B \sim    
1$ and the wave energy is then rapidly damped to heat. Thus, magnetic    
field amplification is not included in this description.     
    
It is via this energy transfer from energetic particles   
that the magnetic field has its largest effect on the acceleration process   
and even small amounts of background heating from the damping of \alf\    
waves can significantly reduce the acceleration efficiency compared to    
the case where only adiabatic heating is included.    
    
As explained in detail in \citet{BE99}, compression ratios $>4$ occur     
in DSA for two reasons.  First, as relativistic particles are produced     
and contribute significantly to the total pressure, their softer     
equation of state makes the shocked plasma more compressible (as     
$\gamma \rightarrow 4/3$, $\Rtot \rightarrow 7$).      
Second, as the highest energy particles escape from the shock, they   
drain away energy flux which must be compensated for by ramping up the   
overall compression ratio to conserve the fluxes.  Just as in   
radiative shocks, this is equivalent to $\gamma \rightarrow 1$ and   
$\Rtot$ can become arbitrarily large.  Compression ratios as large as   
we show here occur in independent steady-state calculations of   
nonlinear DSA which account for particle loss 
\citep[e.g.,][]{Eichler84,JE91,Malkov98,Blasi2002}.  We note that as the   
overall compression ratio increases ($\Rtot > 4$), the subshock
compression ratio, $\Rsub$, which is responsible for heating the gas,
becomes less than the test-particle value ($\Rsub < 4$), causing the
temperature of the shocked gas to drop below \TP\ values.
   
Two important qualifications must be made concerning particle escape 
and the production of compression ratios greater than 7.  First, the 
explicit assumption in the model of Berezhko and Ellison (and the 
others mentioned above) is that steady-state conditions apply and 
particle acceleration is terminated as particles diffuse away from the 
shock. If, instead, the acceleration time becomes comparable to the 
shock age while the diffusion length is still a small fraction of the 
shock radius, the acceleration process may terminate without particle 
escape \citep[e.g.,][]{Drury83}. 
It is less clear what happens in this case, although the work of     
Berezhko and co-workers \citep[e.g.,][]{Berezhko96,BEK96} suggests     
that for the forward shock in SNRs, geometrical factors     
determine $\pmax$ over most of the lifetime. These geometrical     
factors, i.e., the diffusion of particles upstream from the shock, the     
increase in the shock size and upstream volume, the slowing of the     
shock speed, and the adiabatic cooling of the energetic particles,     
produce effects similar to those from escaping particles even though     
particles, in fact, remain in the SNR system \citep[see a comparison     
of the modeling of SN1006 using the simple, steady-state model used     
here with the time-dependent model of Berezhko and co-workers     
in][]{EBB2000}.     
     
The situation is more uncertain for the inward facing, reverse shock     
since the upstream region has a finite volume and particles streaming     
far upstream can conceivably reach the shock on the opposite side of     
the explosion site without being lost. In this case, particles may     
still be lost from spatial effects if they diffuse far enough     
downstream to reach the contact discontinuity. However, large magnetic     
fields are expected at the contact discontinuity due to compression     
and stretching of the field lines by Rayleigh-Taylor instabilities,     
and might act as a magnetic wall confining the particles in the ejecta     
material.     
   
In any case, for either the forward or reverse shock, a precise
determination of the compression ratio in a time-dependent situation
requires a detailed knowledge of the wave-particle interactions in the
self-generated magnetic turbulence responsible for particle diffusion.
This knowledge does not yet exist so approximations remain necessary.
We simply assume that particles at either the forward or reverse shock
leave the system when their acceleration time $> \tSNR$ or their
upstream diffusion length $> \fRad \Rsk$.  In all of the SNR models we
show in this paper, we arbitrarily take $\fRad = 0.05$
and note that  $\pmax$ and $\Rtot$ would increase with increasing
$\fRad$.
   
The second qualification is that the approximations in the Berezhko
and Ellison model assume that $\pmax \gg A \massp c$ and thus, that a
sizable fraction of the total pressure is in \rel\ particles. Despite
this limitation, we show cases where $\pmax \ga 2 A \massp c$ and
warn that our lower $\pmax$ results have a greater intrinsic error
than those with $\pmax \gg A \massp c$.  In fact, from comparisons with
Monte Carlo results (not shown here), we find that the Berezhko and
Ellison model underestimates $\Rtot$ when $\pmax \sim A \massp c$.
   
\section{Efficient Diffusive Shock Acceleration in Weak Magnetic Fields}     

\begin{figure}        
   \centering
   \includegraphics[width=\columnwidth]{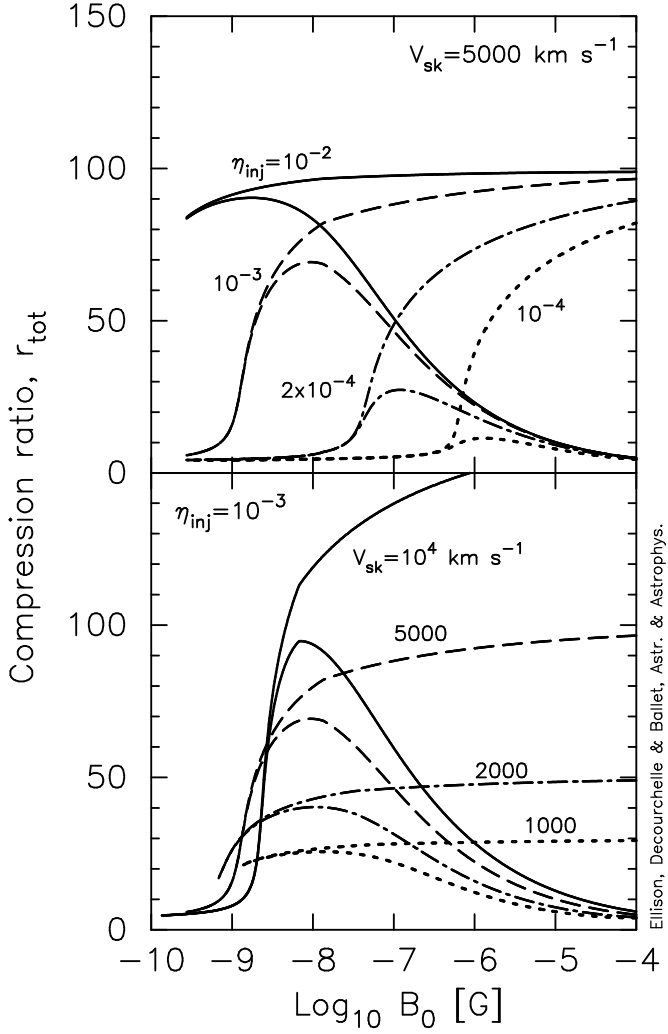}
   \caption{Compression ratio versus unshocked magnetic field for a
   constant shock speed $\Vsk$ and different values of the injection
   efficiency $\etainj$ (top panel), and for a constant injection
   efficiency and different values of the shock velocity (bottom
   panel). For each set of parameters, the upper curve corresponds to
   adiabatic heating in the precursor and the lower one to \alf-wave
   heating in the precursor. In the latter case, $\Rtot$ is
   dramatically reduced as $B_0$ increases.}
   \label{fig_RvB}
\end{figure}     
     
In this section, we investigate how the acceleration efficiency   
depends on the upstream magnetic field $B_0$ for a fixed set of   
parameters; $\Rsk$, $\fRad$, $\tSNR$, and proton number density   
$\np$.  We chose a value of $\tSNR$ large enough to ensure that    
$\pmax$ is determined by diffusive escape. In this case, since we set   
$\etamfp=1$ and $\Rsk$ and $\fRad$ are fixed, $\pmax$ depends   
only on $B_0$ and $\Vsk$:  
\begin{equation}     
\pmax  \simeq 3\, Z e\, \fRad \, \Rsk \, B_0 \Vsk/c     
\ ,     
\end{equation}     
where $Z$ is the charge number or ionization state and $e$ is the 
electronic charge. 
   
Figure~\ref{fig_RvB} shows the variation of the overall shock
compression ratio $\Rtot$ versus magnetic field (top panel) for
different values of the injection parameter $\etainj$ with $\Vsk$ set
to $5000$~\kmps\ and (bottom panel) for different values of the shock
velocity with $\etainj$ set to $10^{-3}$.  In both panels, $\Rsk =
3$~pc, $\tSNR = 1200$~yr, $\np = 0.3$~\pcc, and the plasma is composed
of protons and electrons only.
   
Each value of $\etainj$ has two curves, one with only adiabatic
heating in the precursor and one with \alf-wave heating in the
precursor. The transfer of energy from the energetic particles to the
background gas via \alf\ waves reduces $\Rtot$ dramatically as $B_0$
increases. With \alf-wave heating, very large compression ratios are
obtained only for values of the magnetic field much lower than that
typical of the interstellar medium (i.e., for $B_0 \ll
3\xx{-6}$~G). This compression ratio increases with decreasing
magnetic field up to a maximum, whose position and intensity depend on
the injection parameter, and then decreases to the \TP\ value of 4.
   
In the top panel, the sonic Mach number is fixed and the only relevant
varying parameters are $\pmax$ and the \alf\ Mach number, $\Malf$. As
$B_0$ decreases, $\pmax$ decreases and $\Malf$ increases. For $B_0
\ga 2\xx{-8}$~G, the magnetic field is strong enough to allow a
significant transfer of energetic particle energy through \alf\ waves
to heat, lowering the acceleration efficiency. The greater $B_0$, the
smaller $\Malf$ and the more important this effect becomes, causing
$\Rtot \rightarrow 4$. For $B_0 \la 2\xx{-8}$~G, $\Malf$ is large
enough that it is no longer important, but now $\Pmax$ becomes small
enough that the fraction of pressure in \rel\ particles drops below
that required to maintain a strongly modified shock. As $B_0$
decreases, the transition from a strongly modified shock with $\Rtot
\gg 4$ to an {\it unmodified} one with $\Rtot \sim 4$ occurs and can be
extremely abrupt. As explained in \citet{BE99}, the larger $\etainj$
is, the more difficult it is to have a high-Mach number, unmodified
solution. This is the reason that the maximum value of the compression
ratio increases, and that the position of the maximum shifts towards
lower $B_0$, as $\etainj$ increases.
    
In the bottom panel,  both $\Msonic$ and $\Malf$ vary.   
With \alf-wave heating, as $B_0$ decreases, the curves for $\Vsk =   
1000$ and $2000$~\kmps\ end before the compression ratio drops to 4:   
in these cases, $\pmax$ becomes low enough ($\la 2\, \massp c$) to   
invalidate the approximations of the Berezhko and Ellison model.   
       
As $B_0$ increases and $\Malf$ drops, the damping effects of the
magnetic field increase and cause $\Rtot$ to drop toward 4, regardless
of the sonic Mach number. For values of magnetic field near
$10^{-8}$~G in the \alf-wave heating curves, the compression ratio
peaks strongly as the shock velocity and, therefore, $\Msonic$ increase.
As $B_0$ decreases below $\sim 10^{-8}$~G, the lowering of $\pmax$,
and subsequent reduction of pressure in \rel\ particles, causes the
transition to unmodified solutions regardless of $\Msonic$. The magnetic
field where $\Rtot$ is a maximum shifts slightly toward lower $B_0$ as
$\Msonic$ increases.  Using the Berezhko and Ellison model, an approximate
expression for the magnetic field strength, $B_*$, at the transition
point between unmodified and strongly modified solutions can be
derived in two regimes (for $\pmax \le$ or $> 100\, A \massp c$):
  
\begin{eqnarray}     
B_* & \simeq & \frac{\left(x_1/\sqrt{10} \right)^4}{x_2}
\frac{\Vsk^3}{\etainj^4 c^2} \\ 
& & \mathrm{if~} \pmax > 100\, A \massp c \ , \nonumber
\end{eqnarray}     
or   
\begin{eqnarray}     
B_* & \simeq & \frac{x_1^2}{x_2} \frac{\Vsk}{ \etainj^2} \\    
    & & \mathrm{if~} \pmax < 100\, A \massp c \ , \nonumber   
\end{eqnarray}     
where  
\begin{eqnarray}     
   x_1 & = & \frac{\Rsub-1}{2 \lambda_\mathrm{inj}  \sqrt{\Rsub}} \times \\    
   & & ~\frac{\sqrt{\gamma +1}}{\sqrt{2 \gamma - \frac{\gamma -1}{\Msonic^2}     
   \left(\Rtot/\Rsub\right)^{\gamma + 1}}}  \nonumber  
\end{eqnarray}   
\begin{eqnarray}     
   x_2 & = & \frac{3 \fRad Z e \Rsk}{A \eta_\mathrm{mfp} \massp}   
\ .   
\end{eqnarray}     
At the transition, $\rtot$ can be estimated as $0.65\, \Msonic^{3/4}$ and
$\Rsub = 4 $ \citep{BE99}. The break at $100\, A \massp c$ mirrors the
break in the Berezhko and Ellison model between a three-component
power law when $\pmax \ge 100\, A \massp c$ and a two-component power law
when $\pmax < 100\, A \massp c$.
     
Figure~\ref{fig_RvB} illustrates that even if extremely high
compression ratios are theoretically possible, normal ISM magnetic
field values, i.e., $\BISM \ga 3\xx{-6}$ G, with \alf-wave
heating, are sufficiently high to limit compression ratios to $\Rtot
\la 20$, regardless of how efficient the injection is or how high
the sonic Mach number is. The energy converted to magnetic turbulence
and heat lowers the subshock Mach number and the overall acceleration
efficiency. Also, when $\Malf$ is low, the speed of the magnetic
scattering centers in the fluid (assumed to be \alf\ waves) can become
comparable to the shock speed, lowering the effective difference (for
acceleration) between the downstream and upstream flow
speeds.\footnote{Note that even though we consider cases where the
speed of the scattering centers is high, we neglect second-order Fermi
acceleration.}
The low magnetic fields in expanding supernova ejecta offer a unique   
possibility of seeing non-radiative shocks with $\Rtot \ga 20$.   
     
\section{Particle acceleration at reverse shocks in SNRs}         
 
\subsection{Assumptions and parameters for the CR-hydro simulation}     
 
We model the effects of cosmic-ray acceleration on the evolution 
of a SNR using a one-dimensional, cosmic-ray hydrodynamic (CR-hydro) 
simulation as described in \citet{EDB2004}. 
    
We initialize the CR-hydro simulation at some time $\Tstart$ after the 
explosion with a power-law ejecta density distribution, $\rhoEj 
\propto r^{-n}$ (of index $n = 7$), combined with a constant density 
plateau region at small radii.  The plateau is required to keep the
total ejecta mass finite.  Beyond the ejecta, we assume a uniform ISM
mass density $\rhoISM=5\xx{-25}$~g~cm$^{-3}$ corresponding to a proton
number density of $0.3$~cm$^{-3}$. 
A constant density ISM is more appropriate for a Type Ia
supernova, whereas a Type II supernova is likely to explode in a
pre-SN stellar wind with a $\rho \propto r^{-2}$ density structure. 
The presence of a stellar wind will affect the density
and temperature structure of the shocked ejecta and this, in turn,
will change the
quantitative aspects of particle acceleration at the reverse
shock. Qualitatively, however, the effects we describe for a uniform
ISM will be present in Type II supernovae as well. 

The hydrodynamical simulation only
recognizes the matter density, but the acceleration model depends on
the ion species and is limited to a single species, as we discuss
below.  We assume that the ejecta speed varies linearly with radius
from zero to some maximum speed $\Vmax$.  In the simulation, the
initial maximum radius of the ejecta is set by the maximum ejecta
speed and $\Tstart$.  Thus, the early stages of the simulation will
depend on $\Vmax$ and $\Tstart$. As long as the total kinetic energy
and ejecta mass stay the same, however, the later evolution of the
SNR is independent of both $\Vmax$ and $\Tstart$.

\begin{figure}[!hbtp]              
   \centering 
   \includegraphics[width=\columnwidth]{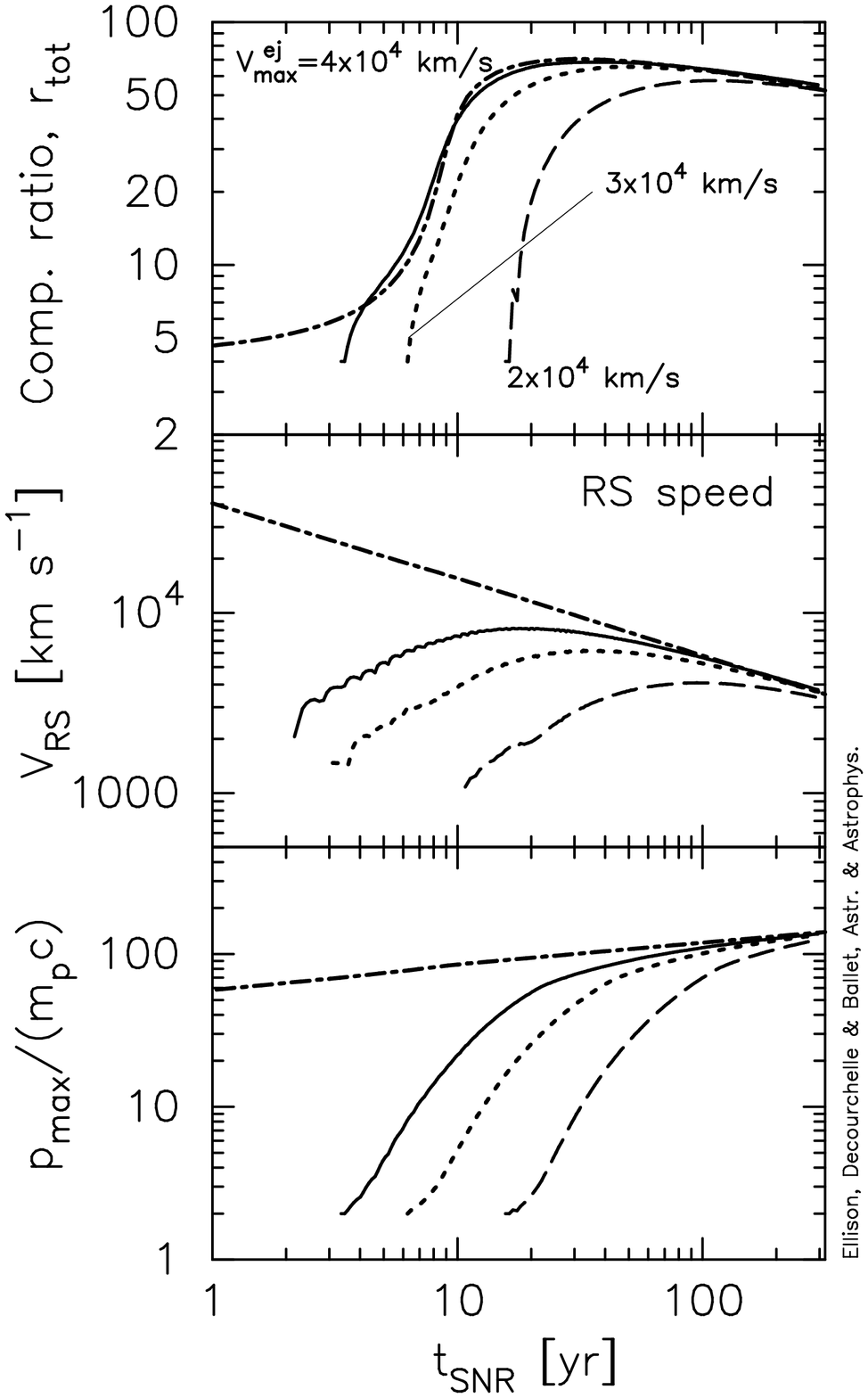}
   \caption{All curves are for the reverse shock and use the same set
   of parameters with a constant magnetic field $\Bej=3\xx{-8}$~G.
   The only variable is the maximum speed of the ejecta $\Vmax$ which
   is $4\xx{4}$~\kmps\ for the solid curves, $3\xx{4}$~\kmps\ for the
   dotted curves, and $2\xx{4}$~\kmps\ for the dashed curves.  For
   comparison, we show the results obtained analytically, where
   $\Vmax$ is not limited, for the same set of parameters using
   modified Chevalier solutions. While all curves converge after
   several decades, the early evolution depends on the different
   hydrodynamic initial conditions. The simulation curves are
   truncated when $\pmax < 2 \massp c$.}
   \label{fig_Vmax}     
\end{figure}     

In Figure~\ref{fig_Vmax} we show how the early evolution depends on
$\Vmax$ for a particular set of SNR parameters.  For all of our other
examples, except those indicated in Fig.~\ref{fig_Vmax}, we take
$\Vmax = 3\xx{4}$~\kmps\ $ \simeq 0.1c$.  As long as $\Tstart$ is
earlier than the time when $\pmax$ becomes greater than $2\, A \massp c$,
as is the case in Fig.~\ref{fig_Vmax}, our results are independent of
$\Tstart$.
   
For comparison, we also show in Fig.~\ref{fig_Vmax} (dot-dashed curve)
the results obtained analytically for the same set of parameters using
modified Chevalier solutions \citep[][]{DEB2000}.  Good agreement is
reached after a few tens of years and, in fact, all of our simulation
results, after a few decades, are independent of our starting
conditions and consistent with analytic solutions, as long as the
self-similar conditions required for the analytic solutions are valid.
 
\subsection{Constant ejecta magnetic field}     

Given the general behavior of DSA in low magnetic fields, we begin our
study of SNRs by assuming a constant ejecta magnetic field upstream
from the reverse shock.
  
For the CR-hydro model, we use the following parameters: supernova
explosion kinetic energy $\EnSN=10^{51}$ erg, ejecta mass
$\Mej=1.4~\Msun$, and $\BISM=3\xx{-6}$~G for the upstream ISM field
(this is $B_0$ for the forward shock). For the acceleration
calculation, we take $\etainj=10^{-3}$ here and in all following
models for both the forward and reverse shocks.\footnote{In an actual
SNR, of course, the injection efficiency might vary with time, vary
over the shock surface, or be different at the forward and reverse
shocks \citep[as in our model of Kepler's SNR;][]{DEB2000}. If the
actual $\etainj$ is less than $10^{-3}$, the nonlinear effects we show
will be less dramatic.}
  
\begin{figure}              
   \centering 
   \includegraphics[width=\columnwidth]{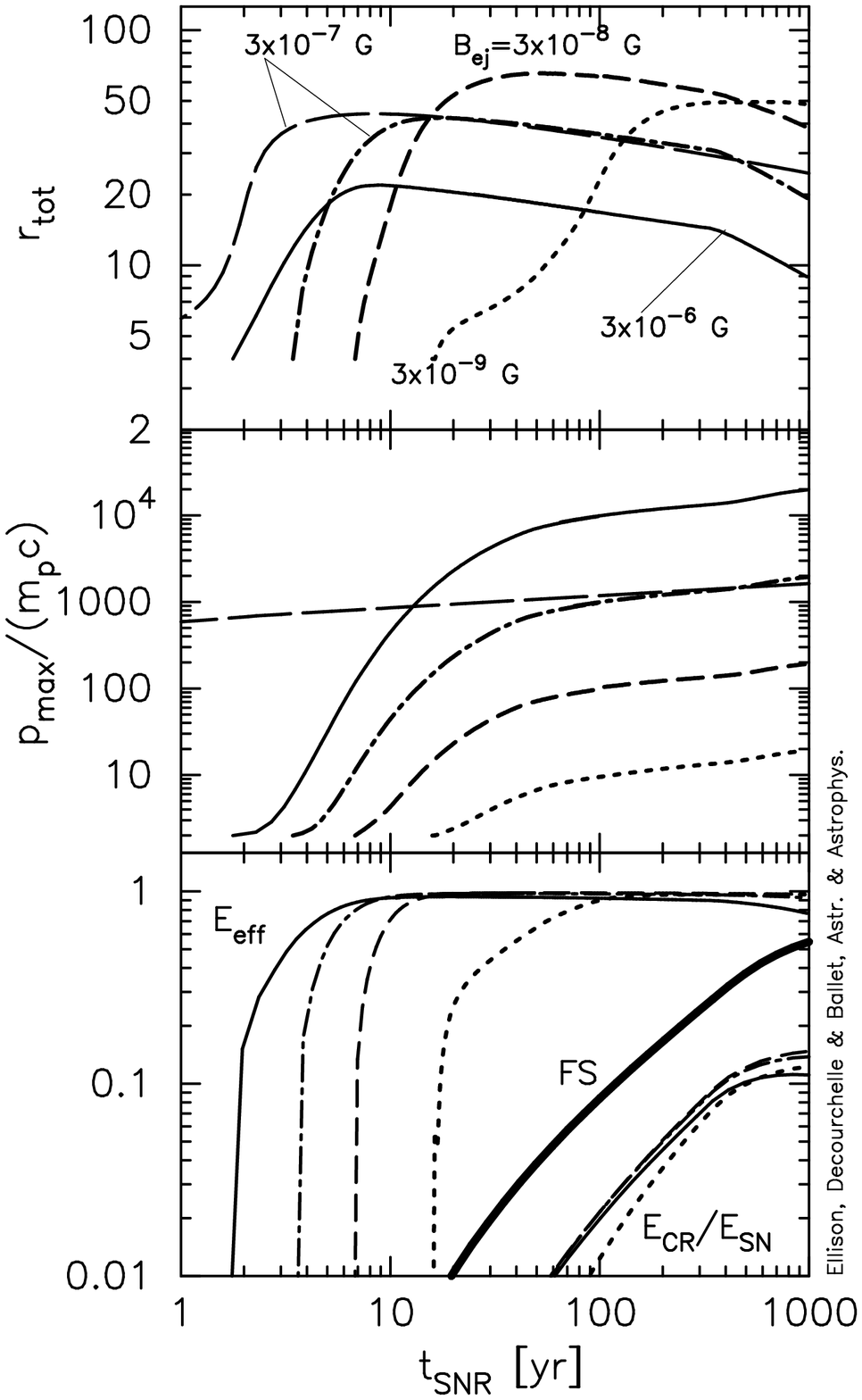} 
   \caption{In each panel, RS results are shown for $\Bej=3\xx{-6}$~G 
   (solid curves), $3\xx{-7}$~G (dot-dashed curves), $3\xx{-8}$~G (dashed
   curves), and $3\xx{-9}$~G (dotted curves).  The long-dashed curves
   in the top and middle panels are analytic results with
   $\Bej=3\xx{-7}$~G.  In the bottom panel, $\Eeff$ is the RS
   acceleration efficiency and $\EnCR/\EnSN$ is the fraction of SN
   explosion energy going into CRs.  The curve labeled `FS' is
   $\EnCR/\EnSN$ for the FS (FS results are insensitive to $\Bej$). In
   all cases, $\BISM=3\xx{-6}$~G and curves are truncated when $\pmax
   < 2 \massp c$.}
   \label{fig_Rtot}     
\end{figure}     

The top panel of Fig.~\ref{fig_Rtot} shows the variation of the
compression ratio with SNR age for different values of the constant
upstream ejecta magnetic field $\Bej$ ($B_0$ for the reverse
shock). For $\Bej \geq 3\xx{-6}$~G, the compression ratio is smaller
than 20 for a SNR age larger than 100~yr.  For $\Bej \leq 3\xx{-7}$~G,
however, the reverse shock compression can be extremely high: $\Rtot
\sim 60$ at $\tSNR \sim 100$~yr for $\Bej=3\xx{-8}$~G (dashed 
curve). The analytical results for $\Bej=3\xx{-7}$~G are shown as
long-dashed curves in the top two panels. As explained for
Fig.~\ref{fig_Vmax}, the early evolution depends on $\Vmax$ but good
agreement is obtained for higher $\Vmax$ or, in any case, after $\sim
10$~yr. The change in the slope in the simulation curves at $\sim
400$~yr corresponds to the passage of the reverse shock from the power-law
envelope into the plateau. Such a transition is not included in the
analytic result which is only valid while the reverse shock remains in
the power-law profile.
 
Lowering $\Bej$ causes the particle gyroradius and acceleration time
to increase so in a shock of a given size and age, $\pmax$ decreases,
as seen in the middle panel of Fig.~\ref{fig_Rtot}. Below some
minimum value, $\Bej$ will be too weak to allow the acceleration of
particles to radio emitting energies. This opens up the possibility
that a range of $\Bej$ may exist greater than the minimum value needed
to produce observable radio emission but less than $\sim 3\xx{-6}$~G
so that the full nonlinear effects of efficient Fermi acceleration of
ions occurs.\footnote{Note that we do not show solutions in
Fig.~\ref{fig_Rtot} for $\pmax < 2\,\massp c$, i.e., below the limit of
validity of the Berezhko and Ellison model.}
   
The efficiency of the DSA process can be extremely high. In the bottom 
panel of Fig.~\ref{fig_Rtot} we show the instantaneous acceleration 
efficiency $\Eeff$, defined as the fraction of incoming energy flux 
(in the shock rest frame) put into \rel\ particles. After a few 
decades, all of the models show $\Eeff \ga 0.9$. The curves in the 
bottom panel on the right show the fraction of the supernova explosion 
energy $\EnSN$ put into \rel\ particles, i.e., $\EnCR/\EnSN$. The 
reverse shocks are able to put $\sim 10$\% of $\EnSN$ into CRs after 
1000\,yr. The forward shocks put $\sim 50$\% of $\EnSN$ into CRs, as 
shown by the heavy-weight solid curve labeled ``FS.''\footnote{As 
noted by \citet{BKV2002}, injection may vary over the surface of the 
SNR and be significantly less where the magnetic field is highly 
oblique. They estimate that to supply the galactic CR population the 
overall efficiency need only be $\sim 20$\% of the maximum values 
obtained by DSA.} 
    
\begin{figure}                      
   \centering 
   \includegraphics[width=\columnwidth]{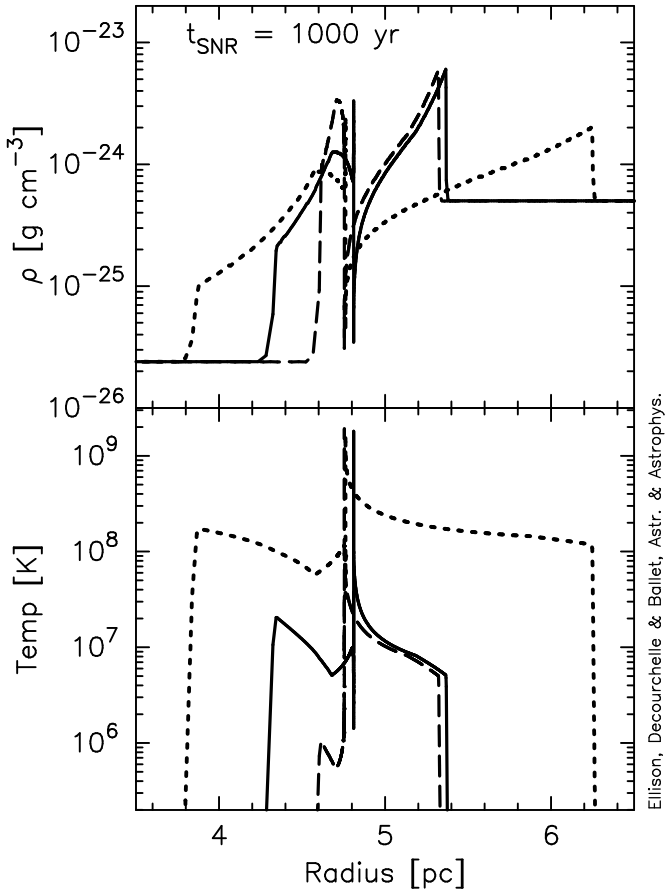}
   \caption{The top panel shows the plasma density vs. radius and the
   bottom panel shows the temperature vs. radius.  In both panels, the
   solid curves include the effects of efficient DSA with a constant
   ejecta magnetic field $\Bej=3\xx{-6}$~G, the dashed curves assume
   $\Bej=3\xx{-8}$~G, and the dotted curves are results with no
   acceleration (i.e., test-particle).  In all cases the results are
   calculated at $\tSNR=1000$ yr with $\BISM= 3\xx{-6}$~G.}
   \label{fig_prof}  
\end{figure}     

In the strong nonlinear regime, radical changes in the structure of   
the interaction region between the forward and reverse shocks occur.   
In Fig.~\ref{fig_prof} we compare nonlinear density and temperature   
profiles with $\Bej=3\xx{-6}$~G (solid curves) and $\Bej=3\xx{-8}$~G   
(dashed curves) against profiles obtained with no acceleration (dotted 
curves). All of the profiles are plotted at $\tSNR=1000$\,yr.  In the   
cases with efficient particle acceleration, the increased compression   
ratios make the interaction region narrower, denser, and cooler than with  
no acceleration.  This is particularly true for the width of the shocked   
ejecta region which, compared to the case with no acceleration, shrinks   
by a factor of $\sim 5$ when $\Bej=3\xx{-8}$~G and where the temperature   
drops from greater than $10^8$ K to $\sim 10^6$ K.   

Note that the density and temperature profiles drop abruptly with
no apparent precursors (the small precursors seen in
Fig.~\ref{fig_prof}, and Fig.~\ref{fig:ejWDprof} below, are numerical
from a finite grid). This is because our current CR-hydro model does
not explicitly include the cosmic-ray precursor in the
hydrodynamics. The precursor effects essential for the nonlinear
acceleration (i.e., shock smoothing and pre-heating) are included in
the particle acceleration model.  For the solid curves in
Fig.~\ref{fig_prof}, the upstream density precursor at the
reverse shock would be $\sim 0.2$\, pc in extent to be consistent with
our assumption that $\fRad \Rsk$ sets the upstream diffusion length of
particles of momentum $\pmax$.  We further note that regardless
of the density precursor, a precursor in thermal X-rays is not expected
because the temperature in the precursor is too low. 
The nonthermal X-ray precursor will be much narrower 
than the density precursor if the electron spectrum is limited by
cooling, as suggested by the narrow width of the filaments.

The changes in structure and evolution of the SNR brought about by
efficient DSA are so large that, even considering the difficulties
projection effects present and other uncertainties, they should be
observable with current techniques.  On the other hand, if radio
emission is unambiguously observed at reverse shocks without such
dramatic structural changes, this would be evidence for either
magnetic field amplification beyond several \muG, or that DSA doesn't
produce large compression ratios as the theory with particle escape
predicts, or that the injection rate is considerably lower than the
value $\etainj=10^{-3}$ we have assumed.
   
\subsection{Non-uniform, diluted ejecta field}     

The structure of the magnetic field in SNR ejecta is clearly more
complex than assumed in the previous section. As the ejecta expands,
the conservation of magnetic flux will cause the field strength to
decrease rapidly and after only a few years, $\Bej$ may fall below the
$3\xx{-8}$ G value used in Fig.~\ref{fig_Rtot}. The highest possible
initial values of the magnetic field are expected for white dwarf (WD)
progenitors of Type Ia supernovae and range between $10^{5}$ and
$10^{9}$~G \citep[e.g.,][]{Liebert95}. Further enhancement of the WD
magnetic field by convection effects could occur prior to explosion
during the phase of quasi-static burning of carbon and could possibly
lead to equipartition between the kinetic energy density and magnetic
field density. If this occurs, fields as high as $10^{10}-10^{11}$~G
might result \citep[e.g.,][]{R-LS98}, providing the upper limit on the
initial magnetic field intensity we consider in the following.
     
After the explosion of the progenitor, the rapid expansion of the
ejecta will dilute the magnetic field.  We obtain expressions for the
magnetic field at time, $t$, in the core (i.e., the plateau region) of
the ejecta $\Bcore$, and in the outer, power-law part of the ejecta
$\Bpl$, as a function of fluid speed $V$, by assuming that the
magnetic field is initially uniform in a constant density progenitor
and then carried passively over during the explosion with the magnetic
flux being conserved during the expansion. Therefore,
\begin{eqnarray}    
\label{eq:BejectaA}    
\Bcore & = & \Bwd \left(\frac{\Vej t}{\Rej}\right)^{-2}\\    
\label{eq:BejectaB}    
\Bpl & = & \Bcore \left(\frac{V}{\Vej}\right)^{-2}    
                        \times \\    
                 & & \left[1+ \frac{3}{n-3}    
                       \left(1-\left(\frac{V}{\Vej}\right)^{3-n}    
                       \right)\right]^{\frac{2}{3}} \nonumber    
\, .    
\end{eqnarray}    
Here, $\Bwd$ is the magnetic field in the white dwarf before the 
explosion, $\Vej$ is the constant fluid velocity at the core-power-law 
transition point, $\Rej$ is the radius within the white dwarf 
enclosing the mass of what will become the core of the ejecta, and $n$ 
is the ejecta density power-law index.  Given $\Vmax$, $\EnSN$ and 
$\Tstart$, $\Vej$ is obtained numerically in the hydro simulation and 
\begin{equation}   
\Rej = \Rwd \left [ \frac{n-3}{n - 3 (\Vmax/\Vej)^{3-n}} \right ]^{1/3},   
\end{equation}   
where $\Rwd$ is the white dwarf radius before the explosion.    
    
Despite specifying white dwarf parameters in the above equations,
similar behavior is expected for the ejecta in Type II supernovae but
starting with a lower magnetic field at a larger radius.
Basically, any pre-SN field will be diluted by the    
expanding ejecta such that $\Bej \propto R^{-2}$. 
     
\begin{figure}                        
   \centering 
   \includegraphics[width=\columnwidth]{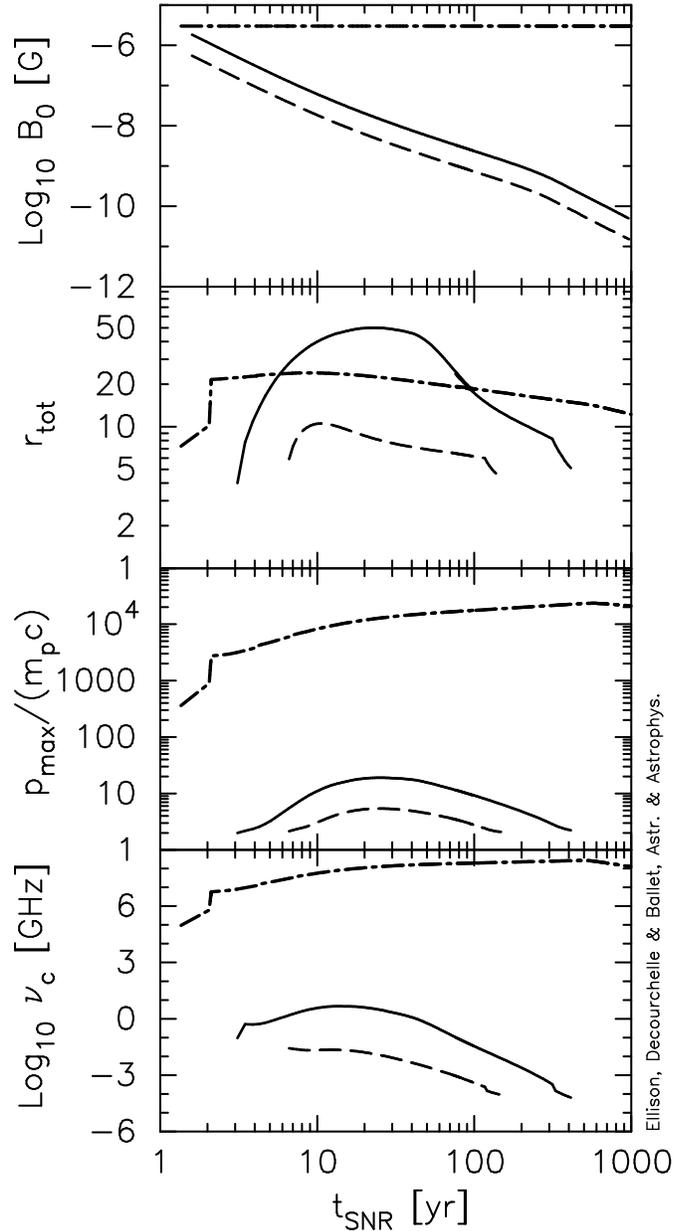} 
   \caption{Results from two models with diluted magnetic fields are  
    shown. In all panels, the solid curves are RS values with  
    $\BWD=10^{11}$~G, the dashed curves are RS values with  
    $\BWD=3\xx{10}$~G, and the dot-dashed curves are FS values, which are  
    insensitive to $\BWD$. For both models, $\BISM = 3\xx{-6}$~G and  
    only values where $\pmax > 2\,\massp c$ are plotted. }
\label{fig:ejWD}     
\end{figure}     

In Fig.~\ref{fig:ejWD}, we show the variations of $\Rtot$, $\pmax$,
and $\nu_c$ versus $\tSNR$ for a diluted field $B_0$, as defined by
eqns.~(\ref{eq:BejectaA}) and (\ref{eq:BejectaB}).  The frequency
$\nu_c$ is the critical frequency \citep[e.g.,][]{RL79} where
\syn\ emission from an electron of $\pmax$ peaks, i.e.,   
\begin{equation}   
\nu_c = 3 \pmax^2 e B / [4 \pi (\masse c)^3]   
\ .   
\end{equation}   
Here $\masse$ is the electron mass and we have taken the sine of the
pitch angle $=1$ for convenience.
   
The solid curves in Fig.~\ref{fig:ejWD} show the extreme case with
$\Bwd=10^{11}$ G.  In this case, the ability of the reverse shock to
produce $\sim 10$~GeV particles lasts until $\tSNR \sim 50$ yr, after
which $B_0$ drops below $\sim 10^{-8}$~G.  The dashed curves show
results for $\Bwd=3\xx{10}$~G.  Despite $\Bwd$ being as high as
$3\xx{10}$~G, few GeV particles are produced at the RS. This is
contrasted by the FS (dot-dashed curves) where $>10^4\, \massp c$
particles are produced after $\sim 100$\,yr.
   
\begin{figure}                        
   \centering 
   \includegraphics[width=\columnwidth]{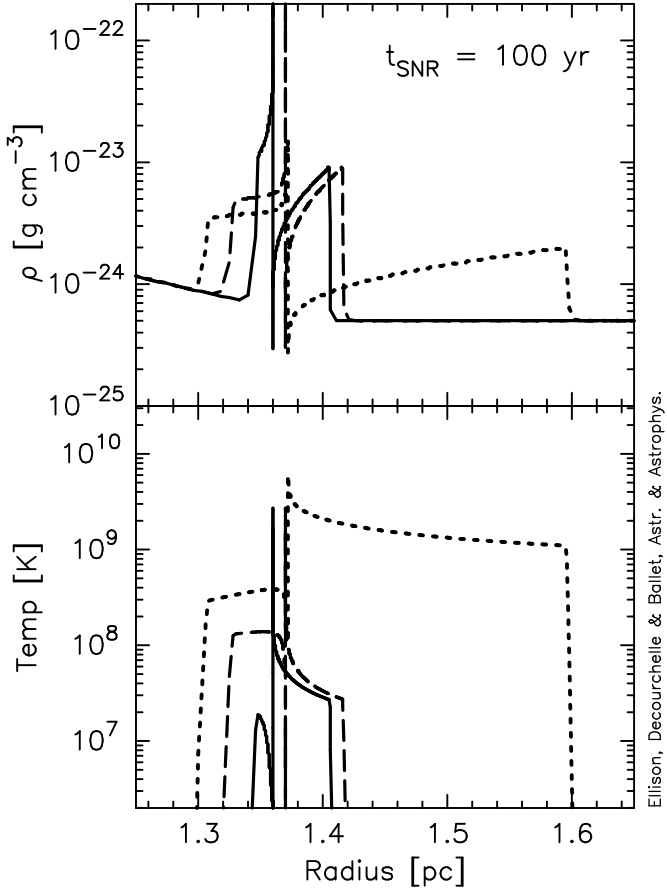}
   \caption{This figure shows the density and temperature profiles
   with diluted ejecta fields. The dotted curves are results with no
   shock acceleration, the solid curves are results with
   $\BWD=10^{11}$~G, and the dashed curves are results with
   $\BWD=3\xx{10}$~G.}
   \label{fig:ejWDprof}
\end{figure}    

Even though the maximum momenta produced in these diluted magnetic 
field examples are low, large structural changes occur early in the 
evolution, as shown in Fig.~\ref{fig:ejWDprof}, where the density and 
temperature profiles are plotted at $\tSNR=100$~yr.  The differences 
between the test-particle case (dotted curves) and the efficient 
acceleration cases (solid and dashed curves) are greater for the 
forward shocks, but are substantial at the reverse shocks.  The 
structure changes at the reverse shocks lessen as $\tSNR$ becomes 
greater than 100\,yr. 

\begin{figure}                        
   \centering 
   \includegraphics[width=\columnwidth]{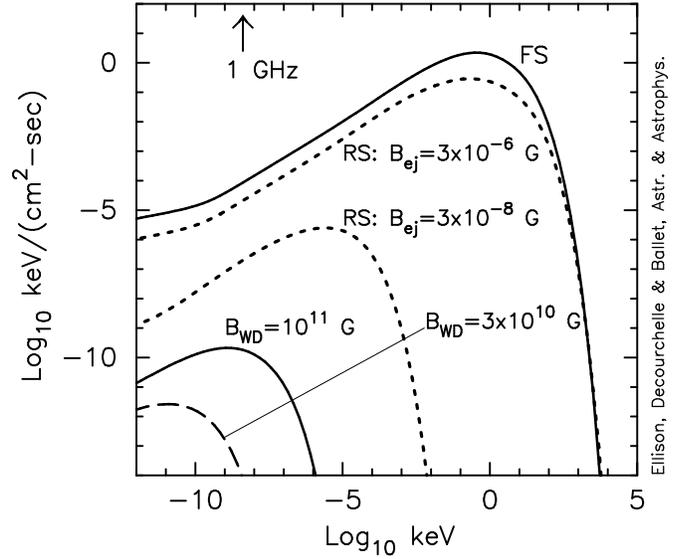}
\caption{Integrated synchrotron emission for constant and diluted 
  ejecta magnetic fields at $\tSNR=1000$~yr.  The upper three curves 
  are calculated assuming a constant field $B_0=3\xx{-6}$~G upstream 
  of the forward shock and constant $\Bej$ as labeled.  The solid and 
  dashed curves are the emission from the reverse shock assuming a 
  diluted white dwarf field of $\BWD=10^{11}$~G and $\BWD=3\xx{10}$~G, 
  respectively.  All curves are calculated for a distance of 2 kpc. }
\label{fig:syn_sn1006}     
\end{figure}

In Fig.~\ref{fig:syn_sn1006} we show the reverse shock \syn\ emission
predicted for the $\BWD=3\xx{10}$ and $10^{11}$~G examples shown in
Fig.~\ref{fig:ejWD}, along with emission from the forward and reverse
shocks where $\Bej$ is held constant at $3\xx{-6}$~G and $3\xx{-8}$~G
as labeled.  For calculating \syn\ emission here and elsewhere, we
assume the electron to proton density ratio at \rel\ energies to be
0.01, similar to that observed for galactic cosmic rays
\citep[see][for discussions of how electrons are treated and \syn\
emission calculated in this model]{BaringEtal99,EBB2000}.
These curves are calculated at $\tSNR=1000$~yr at a distance of 2 kpc, 
typical of SN1006.  At 1~GHz radio frequencies, the reverse shock 
emission from even the most extreme white dwarf case with dilution 
falls more than 5 orders of magnitude below that of the forward 
shock.\footnote{The properties of the forward 
shock are quite insensitive to the ejecta magnetic field so each of 
these four models have similar forward shock \syn\ emission.}
For the constant $\Bej=3\xx{-8}$~G case, the reverse shock emission is
about a factor of 100 below the reverse shock emission with
$\Bej=3\xx{-6}$~G at 1\,GHz. Note that the $3\xx{-8}$~G field is too
weak to produce X-ray \syn\ emission as reported for SNR RCW 86
\citep[][]{rho02}.
   
\begin{figure}                  
   \centering 
   \includegraphics[width=\columnwidth]{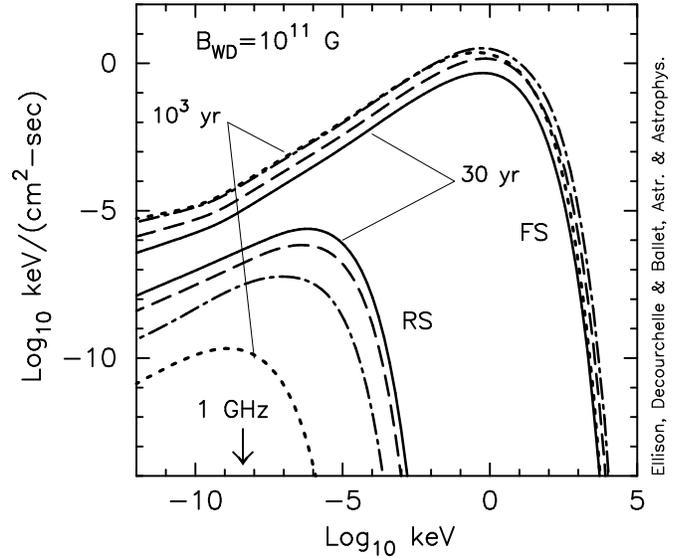}
   \caption{Integrated synchrotron emission for a diluted ejecta
   magnetic field of $\BWD=10^{11}$~G calculated at different ages:
   30~yr (solid curves), 100~yr (dashed curves), 300~yr (dot-dashed
   curves) and $10^3$~yr (dotted curves).  The upper set of curves is
   from the forward shock and the lower set is from the reverse
   shock. Note that the dotted and dot-dashed curves for the FS almost
   overlay each other. The shock and SNR parameters are the same as in
   Fig.~\ref{fig:syn_sn1006}.}
   \label{fig:syn_sn1006_age}  
\end{figure}     

The evolution of the \syn\ emission is shown in 
Fig.~\ref{fig:syn_sn1006_age} for our extreme diluted ejecta magnetic 
field $\Bwd=10^{11}$~G.  The difference in radio emission at 1~GHz 
between the forward and reverse shocks is about a factor of 20 at 30 
years and drops to more than five orders of magnitude at 1000 years, 
as shown in Fig.~\ref{fig:syn_sn1006}. 
    
\subsection{Heavy elements plasma}   

In the examples we have shown so far we considered only acceleration 
in fully ionized hydrogen.  However, ejecta material is expected to be 
composed mainly of heavy elements and, in particular, type Ia 
supernovae are essentially devoid of hydrogen.  There are two reasons 
for the acceleration process to be modified in the case of heavy 
elements. 
  
First, the acceleration time and diffusion length depend on
charge. For \rel\ particles, the time required to accelerate an ion
with charge $q= Ze$ to momentum $p$ is proportional to $1/Z$.
Likewise, the diffusion length of an ion with momentum $p$ is
proportional to $1/Z$ \cite[e.g.,][]{BaringEtal99}. Therefore, for
given shock parameters, $\pmax \propto Z$ regardless of whether
$\pmax$ is determined by a finite shock age or size. A higher $\pmax$
tends to increase the acceleration efficiency if heavy ions are
dominant compared to protons being dominant.
     
Second, a species with mass number $A$ must have momentum $> A\,
\massp c$ to be \rel. Since $\pmax$ only increases as $Z$, this tends
to result in a lower fraction of \rel\ particles, and a $\gamma$
closer to $5/3$, than in the case where protons are dominant. This
tends to decrease the acceleration efficiency. In a mixed plasma
containing light and heavy ions, the modified shock structure
resulting from efficient DSA produces an additional effect whereby
ions with the largest mass to charge ratio are accelerated from
thermal energies most efficiently. This is described in detail in
\citet{EDM97} but is not considered here where we treat only single
component plasmas.
  
\begin{figure}              
   \centering 
   \includegraphics[width=\columnwidth]{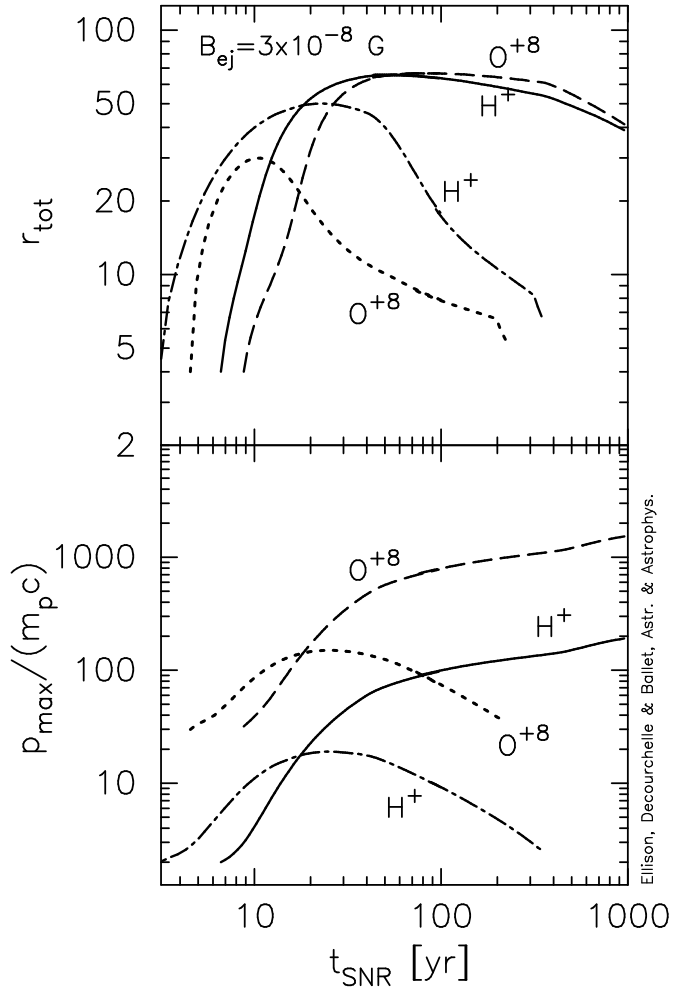}
   \caption{Reverse shock $\Rtot$'s and $\Pmax$'s for material
   containing only protons or fully ionized oxygen, as labeled. The
   ejecta magnetic field is taken to be constant at $\Bej =
   3\xx{-8}$~G in the solid and dashed curves, while a diluted field
   with $\BWD=10^{11}$~G is used for the dot-dashed and dotted
   curves. Only results where $\pmax > 2~A \massp c$ are shown.}   
   \label{fig_AZ}
\end{figure}     

In Fig.~\ref{fig_AZ}, using a constant $\Bej=3\xx{-8}$~G (solid and 
dashed curves) and our extreme white dwarf diluted field 
$\Bwd=10^{11}$~G (dot-dashed and dotted curves), we compare $\Rtot$ 
and $\Pmax$ for a hydrogen plasma and for a fully ionized oxygen 
plasma. All other parameters are the same as those used for the 
examples shown in Fig.~\ref{fig_Rtot}.  For the constant 
$\Bej=3\xx{-8}$~G case, the difference in $\Rtot$ between protons and 
fully ionized oxygen becomes minor after $\sim 50$~yr.  For the 
diluted magnetic field case, however, there are large differences in 
$\Rtot$ between the two species.  The oxygen plasma shows a lower 
maximum $\Rtot$ peaking at an earlier age than the proton 
plasma. Based on this alone, whatever structural and temperature 
changes which occur in the interaction region are expected to be 
greater in hydrogen dominated Type II supernovae envelopes than in 
heavy ion dominated Type Ia supernovae. 
   
\section{Discussion}     

\subsection{Observational evidence for DSA at reverse shocks}    
     
Projection effects and other problems make it difficult to reliably  
associate radio or X-ray nonthermal emission with reverse shocks in  
SNRs. Nevertheless, there have been recent claims for both.  
   
\citet{gotthelf01} have identified the forward and reverse  
shocks in Cassiopeia A and showed that radio and Si emissivity radial  
profiles both show a sharp rise at what they characterize as the  
reverse shock. The sharp rise and fairly sharp falloff in radio  
emissivity as one moves outward from the center of the Cas A SNR is a  
good indication of the local acceleration of relativistic electrons,  
but, as noted, projection effects make the precise determination of  
the reverse shock difficult.  Furthermore, \citet{gotthelf01} chose  
to ignore a secondary Si peak near 80 arcsec inside the main Si  
peak. The main Si peak seems clearly associated with radio emission,  
but the inner secondary peak is not and could alternately mark the  
reverse shock.  
    
Using new VLA observations of Kepler's SNR,     
\citet{delaney02} suggest that two distinct radio structures     
are present. These flat- and steep-spectrum components are partially
decoupled in some areas and the steep-spectrum component tracks the
X-ray emission seen by {\it ROSAT}, which is mainly line emission from
shocked ejecta. They conclude that the flat- and steep-spectrum radio
emission come from the forward and reverse shocks, respectively.
However the steep regions could also mark the interface rather than
the reverse shock. For example, \citet{vink03a} interpreted the (steep)
radio emission from Cas A as arising at the interface.
     
\citet{rho02} claim that the detailed morphology of soft and     
hard X-rays in SNR RCW 86 strongly supports the case for different     
origins for these components. They conclude that the hard X-rays are     
most likely a combination of a \syn\ continuum plus Fe K$\alpha$     
emission from shocked ejecta material. This implies, in their     
estimation, that the reverse shock may be accelerating electrons to     
energies of the order of 50 TeV\,!          
    
While other interpretations are certainly possible, e.g., \rel\ 
electrons produced at the forward shock might travel to the reverse 
shock and brighten in radio in the compressed magnetic field, the 
above observations provide some evidence that reverse shocks in SNRs 
can directly produce \rel\ electrons; GeV in the case of radio and TeV 
in the case of nonthermal X-rays. If this is so, there are a number of 
important consequences both for interpreting a wide range of 
astrophysical sources where DSA is believed to generate \rel\ 
electrons, and for our understanding of the basic functioning of DSA. 
   
\subsection{Implications}   

\subsubsection{Magnetic field amplification}  
     
If \rel\ electrons are accelerated at the reverse shock, the  
production of observable radio emission requires that the magnetic  
field upstream of the reverse shock be amplified by many orders of  
magnitude over values expected in the expanding unshocked ejecta. If  
such field amplification is taking place, it is likely that it is  
directly associated with DSA \citep[i.e.,][]{bell01}.  
  
We speculate that early in the SNR evolution (i.e., $\tSNR \la 10$~yr) 
the unamplified ejecta field may be strong enough to  
start the DSA process even with dilution. Once energetic particles are  
produced, the magnetic field may be amplified by them and maintained  
against dilution at a level where DSA continues.  
  
Shocks are widespread in astrophysics with parameters not widely     
different from those in SNRs. If DSA can amplify magnetic fields in     
SNR shocks, the process should work in other environments as well.      
Since the maximum energy individual particles obtain in DSA scales as   
$B$, widespread $B$-field amplification will lead to a systematic   
increase in the expected maximum proton energy produced by   
astrophysical shocks.   
   
\subsubsection{Extreme nonlinear effects}  

The theory of nonlinear diffusive shock acceleration predicts     
compression ratios far in excess of the test-particle value of $\Rtot     
\simeq 4$ \citep[e.g.,][]{Eichler84,EE84,Malkov98,Bell87,BE99,Blasi2002}.     
For high sonic Mach numbers, $\Msonic^2 > \Malf$, and strongly modified 
steady-state conditions, the compression ratio can be approximated by 
$\Rtot \sim 1.3~\Msonic^{3/4}$ or by $\Rtot \sim 1.5 \Malf^{3/8}$ when 
$\Msonic^2 < \Malf$ \citep[][]{KE86,BE99}.  
 
We have shown that if $\Bej$ falls in a range $3\xx{-8} < \Bej <
3\xx{-6} $~G with typical SNR parameters, $\Rtot \sim 60$ may occur
(see Fig.~\ref{fig_Rtot}); a value considerably larger than is likely
to occur in any other non-radiative astrophysical environment.
Compression ratios this large should produce unmistakable changes in
the SNR structure (see Figs.~\ref{fig_prof} and
\ref{fig:ejWDprof}) and X-ray emission, thermal and nonthermal.  
Confirmation of this prediction would support the premise that DSA is
intrinsically extremely efficient.
    
Unfortunately, shock compression ratios are not directly observable
for remote systems such as SNRs with large Mach numbers \citep[see
sects. 2.2.3.1 and 2.4 in][respectively]{drury2001,raymond2001}.
Shocks in the heliosphere are directly observable but have low Mach
numbers and even here, the direct measurement of compression ratios
requires multiple spacecraft simultaneously sampling the upstream and
downstream plasmas. Until the launch of CLUSTER this was not possible
and the prediction of $\Rtot >4$ has not been clearly demonstrated in
astrophysical shocks, although indirect support does exist for the
Earth bow shock \citep[][]{EMP90}.
     
The large structural changes brought about in SNRs if $\Rtot \gg 4$
offer a unique opportunity to see the effects of extremely efficient
diffusive shock acceleration. In the extreme cases we show here, i.e.,
Fig.~\ref{fig_Rtot}, more than 90\% of the bulk flow energy flux (in
the shock rest frame) is placed in \rel\ ions. Even in cases when
$\Rtot$ is not as extreme, acceleration efficiencies near 50\% are
predicted for both the forward and reverse shocks. Over the lifetime
of a SNR, $\sim 50$\% of $\EnSN$ is predicted to be put into cosmic
rays depending on the average injection rate over the surface of the
SNR \citep[e.g.,][]{Dorfi90,BEK96,BKV2002,EDB2004}. The energy which
goes into \rel\ ions comes out of the bulk thermal plasma and produces
a drastic reduction in the shock temperature.
  
\subsubsection{Cosmic-ray production}  

Direct observations of shocks in the heliosphere and most theories of
DSA show that collisionless shocks put far more energy into ions than
electrons. Thus, even though the presence of \rel\ electrons suggested
our description of strong nonlinear effects, the signature of
nonlinear DSA in the structure and evolution of the radio and X-ray
emitting interaction region between the forward and reverse shocks,
will be evidence for the efficient production of cosmic-ray ions, not
necessarily electrons \citep[see][for a discussion of SN 1006 in this
regard]{BKV2003}. Detection of pion-decay $\gamma$-rays would be a
more direct confirmation that \rel\ ions are produced in SNRs, however
this has not yet been unambiguously done.  It is also clear from
\gamray\ models of young SNRs, that parameters can be chosen where
efficient DSA occurs, but either the \gamray\ flux is below detectable
levels, or detectable TeV $\gamma$-rays are dominated by \IC\ emission
from electrons rather than protons \citep[e.g.,][]{ESG2001}.
  
Modeling the SNR structure offers two advantages. First, if $\Rtot >4$  
is inferred from observations, it is evidence for the efficient  
production of CR ions whether or not a pion-decay feature is observed  
since the observability of pion-decay \gamray's depends on other  
factors besides acceleration efficiency (e.g., ambient density).  
Second, if the structure is inconsistent with efficient DSA and 
$\Rtot \la 4$, this is clear evidence that the efficient shock     
acceleration of ions is {\it not} occurring. Due to the freedom of     
\gamray\ models, the lack of a $\gamma$-ray detection is unlikely ever     
to be able to eliminate the possibility that efficient DSA is
occurring.
  
\subsubsection{X-ray emission}  
 
For efficient particle acceleration, the postshock densities are 
larger and the postshock temperatures smaller than in the \TP\ case 
(see Figs.~\ref{fig_prof} and \ref{fig:ejWDprof}). As a consequence, 
the heating of electrons in the downstream region by Coulomb 
interaction with the population of protons ($T_p \simeq 1836~T_e$ 
without further heating at the shock) will be more efficient than in 
the \TP~case. In the shocked ejecta, the electron temperature can be 
almost equal to that of the protons for an injection of $\simeq 
10^{-2}$, while in the shocked ambient medium the electrons may reach 
up to 30\,\% of the proton temperature \citep{decourchelle2001}. 
 
Another constraint on the efficiency of particle acceleration comes 
from the observation of a strong Fe K-alpha line in the shocked ejecta 
of young SNRs (like Cas A, Kepler and Tycho). In Kepler, for efficient 
particle acceleration at the reverse shock, it was shown that the 
shocked ejecta temperature gets too low to produce the Fe K-line 
\citep{DEB2000}. Line excitation by the nonthermal population can be 
invoked, however the ionization state of iron is expected to be very
low even when taking into account the ionization from a nonthermal
power-law population \citep{porquet2001}.  Thus, the prediction from
efficient DSA that postshock temperatures are low in the RS presents a
problem in SNRs where a strong Fe K-alpha line is observed.
  
\subsubsection{Ionization fraction of ejecta material}   

For DSA acceleration to occur at all, the unshocked ejecta material
must be fully ionized or, at least, have a sizable ionization
fraction. Otherwise the magnetic turbulence necessary to scatter
particles will be damped \citep[e.g.,][]{DDK96}.  The expanding ejecta
will cool rapidly and would be largely neutral unless ionized by some
source. This could possibly be X-ray emission from shock heated gas or
the cosmic rays may contribute to the ionization themselves. If the
ionization fraction is initially large enough for some cosmic-ray
production to occur, these CRs may further ionize the precursor
material, making acceleration more efficient, etc.
 
The Balmer-dominated spectra of nonradiative shocks in a number of SNRs  
\citep[e.g.,][]{ghavamian2001} has been interpreted as charge exchange  
between protons and neutrals. If synchrotron radio and/or X-ray
emission is observed at these shocks as well, which is possibly the
case in the northeastern rim of Tycho, this would indicate that DSA is
at work even in a partially neutral medium.  The reported presence of
reverse shock radio emission suggests that the ejecta material is
ionized, at least in the region upstream of the shock, where DSA is
taking place.

\section{Conclusions}   
     
It has long been believed that forward shocks in SNRs sweep up and  
accelerate ISM ions to produce galactic cosmic rays and to accelerate  
electrons to produce {\it in situ} radio emission. The mechanism most  
likely responsible for this is DSA, which is predicted to be extremely  
efficient, i.e., $>50$\% of the ram kinetic energy may go into  
relativistic ions. At reverse shocks in young SNRs, however,  
straightforward estimates of the diluted ejecta magnetic field from  
the pre-SN white dwarf or massive star show it to be many orders of  
magnitude below that required to either accelerate \rel\ electrons  
from the thermal background, or to produce observable radio  
intensities from background cosmic-ray electrons.  
  
Nevertheless, some recent observations provide evidence for radio
emission associated with reverse shocks in two SNRs: Kepler and Cas A,
and X-ray \syn\ continuum emission in RCW 86. If the reverse shocks in
these remnants are accelerating electrons to GeV or even TeV energies
via DSA, it immediately suggests that the magnetic field at the
reverse shock is orders of magnitude higher than expected to produce
radio and higher still to produce X-ray \syn. If true, the most likely
explanation is that the acceleration process is amplifying the
magnetic field, perhaps as \citet{bell01} have suggested. The
importance of shock acceleration in a wide variety of astrophysical
objects, and the strong dependence of DSA on the magnetic field, make
it critically important to first, verify the difficult reverse shock
observations and second, to explore the ramifications of efficient DSA
at reverse shocks in young SNRs.
  
While radio observations imply magnetic field amplification to values
far larger than the diluted progenitor field, there is, as yet, no way
to precisely determine the value. This opens the possibility that the
ejecta field $\Bej$ is large enough to produce observable radio
emission, but still far lower than normal ISM values.  We have shown
here that, for typical SNR parameters, having $3\xx{-8} < \Bej <
3\xx{-6}$~G results in extremely large nonlinear effects in DSA, i.e.,
compression ratios $\gg 4$ and shocked temperatures $\ll$ than
test-particle temperatures. If these nonlinear effects occur, they
will produce large changes in the structure and evolution of SNRs
which should be observable with existing instruments.  We have
detailed these effects in the remnant hydrodynamics and estimated
\syn\ emission for a limited range of parameters. 
  
Supernova remnants may offer the best known laboratory for studying 
both magnetic-field amplification and DSA. Current and future ground 
and space-based observatories offer high spatial and energy resolution 
of several remnants and provide information that is available nowhere 
else. Shocks in the Heliosphere are accessible to spacecraft and a 
great deal has been learned of their properties. However, the 
point-like nature of heliospheric observations, the low Mach number 
and small size of heliospheric shocks which limits particle 
acceleration, and the unique geometry of the intensely studied Earth 
bow shock, limit what can be learned and transferred to other 
astrophysical systems. The difficult plasma physics has also limited 
the success of analytic investigations, and direct PIC computer 
simulations are decades away from being able to simulate the injection 
and acceleration of a electron-proton plasma to \rel\ energies. 
  
\begin{acknowledgements} 
The authors wish to thank J. Blondin for providing his hydrodynamic
simulation code {\it VH-1} and for other help with this project.  We
are also grateful to the International Space Science Institute (ISSI)
in Bern, Switzerland, where some of this work was done. This work was
supported, in part, by a NSF-CNRS grant (NSF INT-0128883, CNRS-12974)
and by an NASA ATP grant (ATP02-0042-0006).
\end{acknowledgements}

\end{document}